\documentstyle[11pt,cargesepasp,twoside,epsf]{article}
\begin{document}

\def\etal{et al. }
\def\araa{{\it Ann.\ Rev.\ Astron.\ Ap.}}
\def\aplet{{\it Ap.\ Letters}}
\def\aj{{\it Astron.\ J.}}
\def\apj{ApJ}
\def\apjl{{\it ApJ\ (Lett.)}}
\def\apjs{{\it ApJ\ Suppl.}}
\def\aas{{\it Astron.\ Astrophys.\ Suppl.}}
\def\aa{{\it A\&A}}
\def\mnras{{\it MNRAS}}
\def\nature{{\it Nature}}
\def\pasa{{\it Proc.\ Astr.\ Soc.\ Aust.}}
\def\pasp{{\it P.\ A.\ S.\ P.}}
\def\pasj{{\it PASJ}}
\def\pre{{\it Preprint}}
\def\aph{{\it Astro-ph}}
\def\sovlet{{\it Sov. Astron. Lett.}}
\def\adspr{{\it Adv. Space. Res.}}
\def\expas{{\it Experimental Astron.}}
\def\ssr{{\it Space Sci. Rev.}}
\def\ar{{\it Astronomy Reports}}
\def\inpress{in press.}
\def\souspresse{sous presse.}
\def\inprep{in preparation.}
\def\enprep{en pr\'eparation.}
\def\submit{submitted.}
\def\soumis{soumis.}

\def\ap{$\approx$ }
\def\mjysr{MJy/sr}
\def\inu{{I_{\nu}}}
\def\inufit{I_{\nu fit}}
\def\fnu{{F_{\nu}}}
\def\bnu{{B_{\nu}}}
\def\msol{{M$_{\odot}$}}
\def\mic{{{\mu}m}}
\def\cm2{$cm^{-2}$}

\title{Analysis of ProNaOS submillimeter maps in the M42 Orion Nebula ---
 Temperature - spectral index inverse correlation in several regions}
\author{X. Dupac \altaffilmark{1}, M. Giard\\
Centre d'\'Etude Spatiale des Rayonnements (CESR)\\ 
9 av. du colonel Roche, BP4346, F-31028 Toulouse cedex 4, France}

\author{J.-P. Bernard, J.-M. Lamarre, F. Pajot\\
Institut d'Astrophysique Spatiale (IAS)\\
Campus d'Orsay B\^at. 121 \\
15, rue Cl\'emenceau, F-91405 Orsay cedex, France}

\author{I. Ristorcelli, G. Serra \\
CESR}

\author{J.-P. Torre \\
Service d'A\'eronomie du CNRS \\
BP3, F-91371 Verri\`eres-le-Buisson cedex, France}

\altaffiltext{1}{dupac@cesr.fr}

\begin{abstract}
We have mapped with the French submillimeter balloon-borne telescope ProNaOS the thermal emission
of the dust in 
the M42 Orion Nebula. The map obtained is 50' by 40' with an angular resolution of about 3' in the four efficient
wavelengths 200 
$\mic$, 260 $\mic$, 360 $\mic$ and 580 $\mic$. The temperature and index we obtain are highly variable within the studied
region: the 
temperature varies from 12 K to 70 K, and the spectral index from 1.1 to
2.2. The statistical analysis of the
temperature and spectral index 
spatial distribution in this region, as well as in the other regions mapped by ProNaOS, shows an evidence of an inverse correlation between these two
parameters.
We deduce the column densities of gas and the masses. Being invisible in the IRAS 100 $\mic$ survey, some cold clouds
are likely to be the seeds for future star formation activity
going on in the complex.
\end{abstract}


\section{Introduction}

At about 470 pc, the Orion Nebula is one of the nearest massive star formation regions.
The Orion A giant cloud contains the HII region M42, well known as the Great Orion
Nebula.
Behind the H II region is
the OMC-1 molecular cloud, which corresponds to a CO emission peak.
North from OMC-1, one can clearly see an extended "integral-shaped" filament
(ISF below) with two denser
regions called OMC-2 and OMC-3.
This area has
been intensively studied, from optical radiations to radio wavelengths,
often with high angular resolution but restricted to the hot core of the nebula (see for example
the review of Genzel \& Stutzki 1989).

We have studied this region with the submillimeter balloon-borne telescope ProNaOS.
The simultaneous measurements within four bands covering the wavelength range
200 $\mic$ - 800 $\mic$ allows for the first time to constrain both temperature
and spectral index of the dust, and reveals their variation over the whole
complex. A little part of the area mapped here had been already mapped by
Ristorcelli \etal (1998).

\section{Observations\label{obs}}

ProNaOS (PROgramme NAtional d'Observations Submillim\'etriques) is a French
balloon-borne submillimeter experiment, with a 2 m diameter telescope (Buisson \& Durand 1990). The
focal plane instrument SPM (Syst\`eme Photom\'etrique Multibande, see Lamarre \etal 1994) is
composed of a wobbling mirror, providing a beam switching on the sky and four
bolometers cooled at 0.3 K (180-240 $\mic$, 240-340
$\mic$, 340-540 $\mic$, 540-1200 $\mic$). The beam diameters are 2' in bands
1 and 2, 2.5' in band 3 and 3.5' in band 4.
The true
sky is optimally restored from the time-ordered data by using a Wiener linear
inversion method.
The data we present here were obtained during the second flight of ProNaOS in
september 1996, at Fort-Sumner, New Mexico.

\section{Results\label{res}}

We present in Fig. 1 the images obtained in the extreme photometric bands of
ProNaOS-SPM.

\begin{figure}[ht]
\epsfbox{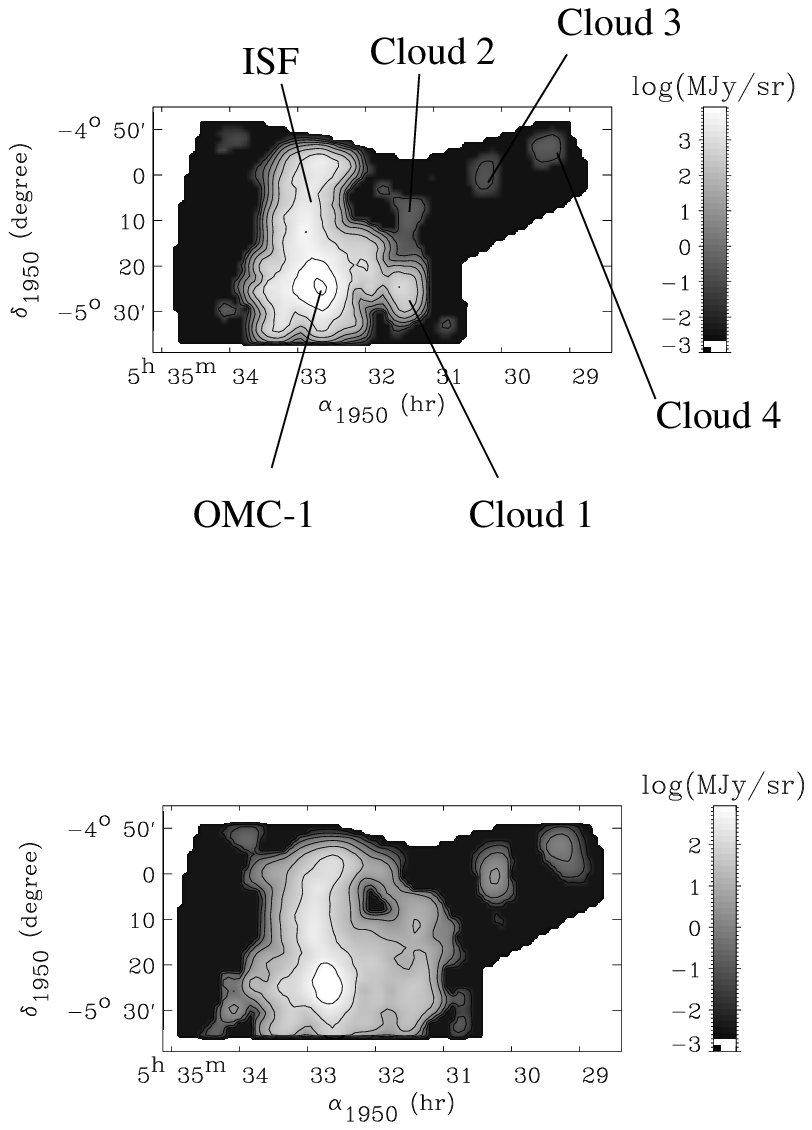}
\caption[]{ProNaOS maps in bands 1 (up) and 4 (bottom). The external contour is the limit of the mapped area.}
\end{figure}

The brightest observed area is the molecular cloud OMC-1.
North from it is clearly visible the integral-shaped filament.
One can see west from the central core a weaker cloud that we call Cloud 1. A weaker emission can be seen around this cloud, that
is linked to the Cloud 2 northwards.
The Clouds 2, 3, 4 can be seen on the west of the maps on Fig. 1, from
east to west. The Cloud 2 is a weak intensity condensation discovered during the first flight of ProNaOS
(Ristorcelli \etal 1998).
The Clouds 3 and 4 are two extremely weak condensations, discovered with this
ProNaOS flight.

\section{Analysis\label{ana}}

\subsection{Dust temperatures and spectral indexes\label{fit}}

We average the intensity of bands 1, 2, 3
with an adequate profile in order to obtain for each band the fwhm of the
fourth band (3.5'). We model the emission of the grains with a modified blackbody law:

$\inufit(\lambda,T,n)=C . \bnu(\lambda,T) . \lambda^{-\beta}$

where $\lambda$ is the wavelength, C a constant, T the temperature of the grains, $\beta$ the spectral
index and $\bnu$ the Planck function.

The three parameters C, T and $\beta$, are adjusted with a least square fit.
The best parameters for each region selected are presented in Table 1.
The dust temperature is highly variable inside and in the neighborhood of this
molecular complex: from {\bf 12 K} to {\bf 70 K}; and the spectral index changes much too:
from {\bf 1.1} to {\bf 2.2}.

\begin{table}
\caption[]{\label{table: results}
Temperature and spectral index from the fit of the spectra, optical
depth. The error bars are given for 68 \% confidence interval. The
approximative diameters of the regions are given in arcminutes.}

\begin{flushleft}
\begin{tabular}{lllll}
\hline
   & T (K) & $\beta$ & ${\tau_{\nu}}\over{10^{-3}}$ & ${\tau_{\nu}}\over{10^{-3}}$\\
   & & & 200 $\mic$ & 580 $\mic$\\

\hline

OMC-1(3.6') & 66.1 $^{+10.2} _{-9.4}$ &
1.13 $^{+0.06} _{-0.09}$ & 47 $\pm $ 6 & 10.4 $\pm $ 0.8\\
\hline

ISF & & & & \\
(south) 5.4' & 22.4 $^{+2.1} _{-2.0}$ & 1.71 $^{+0.12} _{-0.19}$ & 5.3 $\pm $ 1.0 & 1.3 $\pm $ 0.1\\
\hline

ISF & & & & \\
(north) 4.2'  & 25.2 $^{+2.5} _{-2.5}$ & 1.68 $^{+0.13}
_{-0.17}$ & 15 $\pm $ 3 & 1.9 $\pm $ 0.2\\
\hline

Cloud1 (3.5') & 17.0 $^{+3.4} _{-2.5}$ & 2.21 $^{+0.23}
_{-0.48}$ & 4.0 $\pm $ 3.2 & 0.76 $\pm $ 0.23\\
\hline

Cloud2 (3.5') & 11.8 $^{+0.6} _{-0.7}$ & 2.20
$^{+0.15} _{-0.18}$ & 4.2 $\pm $ 1.5 & 0.43 $\pm $ 0.07 \\ 
\hline

Cloud3 (3.5') & 13.3 $^{+2.6} _{-2.1}$ &
1.98 $^{+0.39} _{-0.79}$ & 1.4 $<4.2 (3 \sigma)$ & 0.19 $\pm $ 0.07\\
\hline

Cloud4 (5.2') & 16.9 $^{+7.3} _{-4.1}$ & 1.91 $^{+0.00}
_{-1.29}$ & 0.5 $<2.1 (3 \sigma)$ & 0.12 $\pm $ 0.06\\
\hline

\end{tabular}
\end{flushleft}
\end{table}

\subsection{Column densities\label{model}}

Using the dust 100 $\mic$ opacity from D\'esert \etal (1990), and assuming that the spectral index does not change in the
ProNaOS spectral range, we make a simple self-consistent
model that allows us to estimate the column density $N_{H}$, as a function of
the spectral intensity and the spectral index. There is then a proportionality relation between the constant C and $N_{H}$:

$N_{H}$ = 1.67 $10^{24}$ . C . (100  $\mic)^{-\beta}$

with $N_{H}$ in protons/cm$^{2}$ and C in $\mic^{\beta}$.

We make another simple model using the 100 $\mic$ opacity of
Ossenkopf \& Henning (1994) model specific to protostellar cores.
The proportionality relation is then:

$N_{H}$ = 6.02 $10^{23}$ . C . (100  $\mic)^{-\beta}$

with $N_{H}$ in protons/cm$^{2}$ and C in $\mic^{\beta}$.

We compare our column density results to those derived from observations
of the rotational transition $J = 1-0$ of $^{13}CO$, made by Nagahama \etal
(1998).
We transform the observed $W_{^{13}CO}$ into the column density in
cm$^{-2}$ by using the correspondence established by Bally \etal (1991) in the
Orion Nebula:
$N_{H}/W_{^{13}CO}$ = 2 10$^{21}$ cm$^{-2}$/(K km/s).
As we can see in Table 2, there is a rather good agreement between N$_{H}$ estimated by the model from
D\'esert \etal and N$_{H}$ from $^{13}$CO data, except for the cold clouds 3
and 4. That may be explained by the fact that some cold condensations are the
site of formation of molecular ice mantles on the grains, and of coagulation
of grains. These processes are taken into account by the model of Ossenkopf \&
Henning. The estimations of N$_{H}$ from this model are in rather good agreement with
$^{13}$CO data for Clouds 3 and 4.

\begin{table}
\caption[]{\label{table:interpretation}
Column densities estimated from the opacities of D\'esert \etal (90), Ossenkopf \&
H. (94), and from the $^{13}$CO data of Nagahama \etal (98), mass, Jeans mass.}

\begin{flushleft}
\begin{tabular}{lllllll}
\hline

&N$_{H}$ \tiny{D\'esert} & N$_{H}$ \tiny{Ossenkopf} & N$_{H}$ \tiny{$^{13}$CO}
& \tiny{Mass D\'es.} & \tiny{Mass Oss.} & \tiny{Jeans m.} \\

&$10^{20} cm^{-2}$ &$10^{20} cm^{-2}$ & $10^{20} cm^{-2}$ & (\msol)
& (\msol) & (\msol)\\
\hline

OMC-1(3.6') & 1400 & ------ & 1080 & 212 & --- & ---\\
\hline

ISF & & & & &\\
(south) 5.4' & 245 & ------ & 540 & 83 & --- & ---\\
\hline

ISF & & & & &\\
(north) 4.2'  & 525 & ------ & 680 & 108 & --- & ---\\
\hline

Cloud1 (3.5') & 320 & 115 & 350 & 42 & 15 & 12.6\\
\hline

Cloud2 (3.5')  & 320 & 120 & 200 & 46 & 17 & 8.7\\ 
\hline

Cloud3 (3.5')  & 95 & 34 & 40 & 9.9 & 3.6 & 8.4\\
\hline

Cloud4 (5.2')  & 33 & 12 & 10 & 10.3 & 3.7 & 18.5\\
\hline

\end{tabular}
\end{flushleft}
\end{table}

We also show in Table 2 the mass of each region. In order to have a first evaluation of the gravitational stability of the
clouds, we compare their masses to their Jeans masses.
We see a trend that the closer to the complex the cold clouds are, the more
unstable they seem to be.

\subsection{Temperature - index inverse correlation\label{corr}}

The correlation coefficient between T and $\beta$ has been computed globally for each
studied region of ProNaOS second flight.

$C={\sum{(T_{i}-\bar T) . (\beta_{i}-\bar \beta)}\over\sqrt{\sum{(T_{i}-\bar T)^{2}}
  . \sum{(\beta_{i}-\bar \beta)^{2}}}}$

The correlation coefficients are {\bf -0.92} for Orion, {\bf -0.78} for M17,
{\bf -0.89} for $\rho$ Ophiuci.
From our simulations, this inverse correlation has to be an
intrinsic physical property of the grains.

\section{Conclusion}

Our study shows a large distribution of temperatures and spectral indexes all
around the M42 Orion Nebula. The temperature varies from 12 K to 70 K, and the spectral index from 1.1 to
2.2. The discovery of two new cold clouds (Clouds 3 and 4) confirms that the
existence of cold condensations in such regions is not unusual.
The statistical analysis of the temperature and spectral index spatial
distribution shows an evidence for an inverse correlation between these two
parameters.
This effect is not well explained yet, especially in the submillimeter
spectral range for cold grains ($<$ 20 K). It has been shown to occur in the
laboratory for warm grains
by Mennella et {\it al.}, and for cold grains in the millimeter by Agladze et {\it al.}

We see a trend that the closer to the
complex the cold clouds are, the more unstable they are. The history of star formation around OMC-1 shows
that there has already been 3 to 4 successive bursts of star formation
in this region with the embedded cluster responsible for the BN/KL object
being the latest. The clouds that we observe close to the active region may
thus be the seeds of the next generation of stars. This should be sustained by more observations, particularly of the possible embedded protostars.

\end{document}